\documentclass[a4paper,11pt]{article}
\usepackage{jinstpub} 
\usepackage{lineno}
\usepackage{acronym}
\usepackage{hyperref}

\title{\boldmath A Cloud-native Agile approach to cyber platform prototyping and integration for astronomy: the ENGAGE SKA case}

\author[a,e]{Domingos Barbosa\footnote{Corresponding author.},}
\author[f]{Diogo Regateiro}
\author[b,g]{João Paulo Barraca}
\author[j]{Dzianis Bartashevich}
\author[c]{Marco Bartolini}
\author[d]{Matteo di Carlo}
\author[c]{Piers Harding}
\author[e]{Dalmiro Maia}
\author[e]{Bruno Morgado}
\author[k]{Domingos Nunes}
\author[f]{Bruno Ribeiro}
\author[f,h]{Bruno Coelho}
\author[i]{Valério Ribeiro}
\author[e]{Allan K. de Almeida Jr}
\author[e,h]{Timothée Vaillant}
\author[c]{U\u{g}ur Yilmaz}

\affiliation[a]{HPC Chair, University of Évora, R. Romão Ramalho 59, 7000 Évora, Portugal}
\affiliation[b]{DETI, University of Aveiro, Campus Universitário de Santiago, 3810-193 Aveiro, Portugal}
\affiliation[c]{SKA Observatory, Jodrell Bank, Lower Withington, Macclesfield Cheshire, SK11 9FT, United Kingdom}
\affiliation[d]{Osservatorio Astronomico D’Abruzzo, INAF, Via Mentore Maggini, s.n.c.
64100 Teramo, Italy}
\affiliation[e]{Observatório Astronómico Manuel de Barros, CICGE, DGAOT, Faculdade de Ciências da universidade do Porto, Alameda do Monte da Virgem, 4430-146 Vila Nova de Gaia, Portugal}
\affiliation[f]{ATLAR Innovation, Rua Rangel de Lima, Edifício Multiusos, 3320-229 Pampilhosa da Serra, Portugal}
\affiliation[g]{Instituto de Telecomunicações, Universidade de Aveiro, Campus Universitário de Santiago, 3810-193 Aveiro, Portugal}
\affiliation[h]{CFisUC, Departamento de F\'{\i}sica, Universidade de Coimbra, 3004-516 Coimbra, Portugal}
\affiliation[i]{KLM, Data \& Technology - Platform Ground, PO Box 7700, 1117 ZL Schiphol, The Netherlands}
\affiliation[j]{192.com LTD, London, United Kingdom}
\affiliation[k]{Mercedes-Benz.io, Avenida Dom João ll, 41, 5 Piso 1990-096 Portugal}

\emailAdd{domsilbar@gmail.com}

\acrodef{safe}[SAFe]{Scaled Agile Framework}
\acrodef{mvp}[MVP]{Minimal Viable Product}
\acrodef{ska}[SKA]{Square Kilometre Array}
\acrodef{skampi}[SKAMPI]{SKA MVP Prototype Integration}
\acrodef{skao}[SKAO]{Square Kilometre Array Observatory}
\acrodef{gpu}[GPU]{Graphics Processing Unit}
\acrodef{oci}[OCI]{Open Container Initiative}
\acrodef{bdd}[BDD]{Behaviour-Driven Development}
\acrodef{ssd}[SSD]{Solid-State Drive}
\acrodef{tm}[TM]{Telescope Manager}
\acrodef{csp}[CSP]{Central Signal Processor}
\acrodef{sdp}[SDP]{Science Data Processor}
\acrodef{vpn}[VPN]{Virtual Private Network}
\acrodef{ups}[UPS]{Uninterruptible Power supply}

\abstract{The Square Kilometre Array (SKA) Observatory is gearing up the formal construction of its two radio interferometers in Australia and South Africa after the end of design and pre-construction phases.  Agile methodologies, the Cloud native Computing technologies and the DevOps software ideas are influencing the design of compute infrastructures that will be key to reduce the operational costs of SKA while improving the control and monitoring of the SKA antennas and ancillary systems, Correlators, HPC facilities or related data centre tiered systems. These tools will likely include advanced power metering technologies and efficient distribution automation and Network Operation Centres (NOC). SKA will become the world's largest radio telescope and is expected to achieve its first science by 2026. To cope with this dimension and complexity, a key part of this distributed Observatory is the overall software control and monitoring system embodied in the Observatory Management and Control (OMC) and the Services Teams that requires specialized Agile Teams to assist in software and cyber infrastructure building using an Agile development environment that includes test automation, Continuous Integration, and Continuous Deployment. To manage such a large and distributed machine, the Agile approach was adopted for the core software package of the SKA Telescope aimed at scheduling observations, controlling their execution, monitoring the telescope status and ensuring scalability and reliability. Here, we report on the ENGAGE SKA ciberinfrastructure prototyping support to the SKA Agile Software Development Life Cycle (SDLC).}


\begin{document}
\maketitle
\flushbottom

\section{Introduction}
\label{sect:intro}

The international Square Kilometre Array Observatory (SKAO) started the construction of its planned two radio interferometers in the Southern Hemisphere. These have largely ended their design phase in 2021 and are gearing up for the period of formal construction expected to last until 2028 with first science results to happen by 2026, having already achieved First Light on January 25th 2024.  The SKA is rightly perceived as an iconic, revolutionary new radio telescope and an example for global cooperation in many ‘frontier’ domains of the 21st century\cite{grainge}. The first part of the project, the SKA1, will comprise 130.000 low frequency antennas (50 MHz to 350 MHz) to be deployed in Australia (SKA-Low) and approximately 200 mid frequency antennas (350 MHz to 15.5 GHz) in South Africa (SKA1-Mid). The raw data rate produced will exceed  ~10 Tb/s, requiring a computing power of 100 Pflop/s and an archiving capacity of hundreds of PB/year. The key requirements for the project are a very demanding availability of 99.9\% for its operations, computing scalability and scientific outcome reproducibility. To achieve the enormous potential that the SKAO offers in terms of increasing our understanding of the Universe, exploring technologies for communication and innovation, and incorporating viable green energy supply, the SKA project evolved to maturity through a long period of design and prototyping following rigorous System Engineering principles. The goal is to have a single observatory entity, that will construct and operate two SKA telescopes in two continents (SKA-Mid - South Africa and SKA-Low - Australia), and the Headquarters and Central Operations in United Kingdom. A key part of this distributed Observatory is the overall software control and monitoring system embodied in the Observatory Management and Control (OMC) requiring a Service layer adopted from industry standards of practice. At its heart and considering the SKA geographically distributed nature, the technologies and products developed and selected for Construction were devised to minimize Capital Costs (CAPEX) and Operational Costs (OPEX) while providing high availability and reliability, good maintainability and a solid upgrade path. 

The SKA Construction Schedule is planned through the phased roll-out of dishes and antennas by expanding the interferometer Array Assemblies (AAs), milestones providing an end-to-end telescope system with pre-defined functionality\cite{ska_deliveryplan}.  Of course, construction is a continuous process with the addition of dishes and antennas after each AA release (see table\ref{tab:array_assemblies}), provided sufficient resources for power, compute and network are available on site to accommodate the increasing data rates.  

\begin{table}[ht]

\caption{SKA Construction Schedule - Array Assemblies (AAs). \\AA* is a planned milestone to which the Staged Delivery program may have a planned pause before further funding commitment for the full baseline design (AA4). Last Update - Novembre 2024; from:\\ \url{https://www.skao.int/en/science-users/599/scientific-timeline}}
\label{tab:array_assemblies}
\begin{center}       
\begin{tabular}{l|c|c|c|c} 
\hline
\rule[-1ex]{0pt}{3.5ex}  Milestone Event &Mid Dishes  &Mid Date &Low Stations & Low Date\\
\hline
\rule[-1ex]{0pt}{3.5ex}  AA0.5	& 4 (4 SKA + 0 MeerKAT) &	2026 Q2 &	4	&2024 Q4 \\
\rule[-1ex]{0pt}{3.5ex}  AA1	& 8 (8+0)	& 2027 Q1&	16	&2026 Q1 \\
\rule[-1ex]{0pt}{3.5ex}  AA2 & 64 (64+0)	& 2027 Q4	& 68&	2026 Q4 \\
\rule[-1ex]{0pt}{3.5ex}  {\bf Science Verification}  & & & & \\
\rule[-1ex]{0pt}{3.5ex}  {\bf begins 2027} & & & & \\
\rule[-1ex]{0pt}{3.5ex}  AA*	& 144 (80+64)	&2028 Q3	&307	&2028 Q1 \\
\rule[-1ex]{0pt}{3.5ex} {\bf Early Operations 
 } & & & & \\
 \rule[-1ex]{0pt}{3.5ex} {\bf  begin 2029 (shared risk)} & & & & \\
\rule[-1ex]{0pt}{3.5ex} AA4 	& 197(133+64)&	TBD	&512&	TBD\\
\rule[-1ex]{0pt}{3.5ex}  (Full Design Baseline)	& &		& &	\\
\hline 
\end{tabular}
\end{center}

\end{table} 

Although science commissioning  starts as soon as the first dish/station is available on site, there is a particular interest in these early AA milestones\cite{ska_deliveryplan}:
\begin{itemize}
    \item AA0.5 is the first test array for interferometry, using prototype dishes and receivers for Mid.  This is an engineering array, used to discover system level issues early and develop procedures (e.g., pointing, tracking, holography).
    \item AA2 is the start of science verification, still without the integration of MeerKAT dishes.  The performance of SKA at AA2 is at least as good as existing facilities in many aspects, and usually better (primarily in sensitivity). Observations with AA2 will be used to ensure the system meets the needs of science users (verification of science requirements, testing of observing modes). 
\end{itemize}

The Construction schedule with its AA expansion will require Agile software deployment to cope with the integration of the dishes and the ever-growing of the required data processing facilities. The Software developments and applications are aligned with the best practices advised by the industry, encompassed through principles set by The Software Engineering Institute at Carnegie Mellon University. Throughout this long design process with the utilization of pathfinders and precursors for science and technology anticipation, we can point to the following (non-exhaustive) SKA key major drivers for innovation from ICT and sensor technology to green energy with foreseen contributions to human capital development and employment. In this aspect, Cloud Computing technologies and tools have emerged as a promising Green ICT strategy, which can be exploited by Big Data Centres and Science Organizations \cite{2012Msngr.148...39W,2016SPIE.9910E..0JB}, thus addressing the management and power concerns of large scale science infrastructures. 

Hence, the concepts of Infrastructure as a Service (IaaS), Platform as a Service (PaaS), Software as a Service (SaaS) are key principles for the SKA operation and management platforms and can provide abstraction from the physical compute infrastructures and position data centre operators to trim operational costs while also contributing to reduce the data centres' carbon emission footprint. These will be key components of the Workflows as a Service (WaaS), an architecture ingredient to the science delivery services to be provided by the future SKA Regional Centre network (SRCnet)\cite{2024ASPC..535..399S}, already explored through the exercise of SKA Data Challenges\cite{bonaldi}.  Furthermore, software development sourcing on the DevOps ideas from the Telco/IT sectors promotes agile resource management, automating the process of software delivery and infrastructure through microservice architectural delivery with high modularity. DevOps practices ensure a set of Architecturally Significant Requirements (ASRs) such as deployability, modifiability, testability, and monitorability. These ASRs require a high priority, allowing the architecture of an individual service to emerge through continuous refactoring, hence reducing the need for a big upfront design, reconfiguration of physical infrastructure underneath and reducing the time to market introduction of well-developed software services via frequent software releases early and continuously that prevent premature optimization through adoption of continuous short-term focus. Furthermore, this is coupled to the introduction of Earning Value approaches within organizations to achieve the success of a Project, which is expected by any client from the bidding process to project delivery that emphasize a more comprehensive and typically longer range plan. This approach tries to mitigate problems and avoid frequent replans and reprioritizations since this may be indicative of software delivery or team management problems\cite{2016SPIE.9911E..0NK}.

\section{State of the Art}
\label{sect:sota}

 The maturing of large scale science and industry projects is structured through several phases from design to deployment  and requires a complex set of procedures often lasting several years. Most physics and astronomy projects have now chosen Agile methodologies for their iterative software-development used by project teams for monitoring and control systems or data management systems. The SKA Agile Software Development Life Cycle (SDLC) follows a structured series of stages that a product goes through as it moves from beginning to delivery. It contains six phases: concept, inception, iteration, release, maintenance, and retirement.

A major driver for software development is the cloud-native approach where software developers break functionalities into smaller microservices that work independently and take minimal computing resources to run. Cloud native ecosystem is the software approach of building, deploying, and managing modern applications in cloud computing environments. SKAO has adopted Scaled agile Framework  (SAFe\textsuperscript{\textregistered}), a set of organizational and workflow patterns for implementing agile practices at an enterprise scale. This framework is a body of knowledge that includes structured guidance on roles and responsibilities, how to plan and manage the work, and values to uphold. It was chosen to help Agile Release Train (ART) teams to scale up SKA software development. 

SKA Agile teams have standardised on GitLab\cite{Gitlab} as the social coding platform and orchestrator for the SDLC (through the runner integration), as well as using custom integrations with the GitLab and Nexus\cite{Nexus} repository APIs to automate core software quality controls and security of the software supply chain. GitLab binds together: Developers; Quality framework and standards; Integration \& Delivery; Articulates the Life Cycle processes; and finally Audit and traceability.

GitLab is also essential to help address potential downsides from Agile processes : these require careful communication management, especially in larger teams, to ensure consistent messaging and coordination. Therefore SKAO implemented an entreprise-oriented culture to retain improved project management, consistent output quality and consider Risk mitigation so that each phase includes steps to identify and address risks and reduce probability of costly errors. 

Modern companies want to build highly scalable, flexible, and resilient applications that they can update quickly to meet customer demands. Cloud Native exploits the advantages of the Cloud Computing delivery model: provides Platform as a Service (PaaS) layered on top of an Infrastructure as a Service (IaaS); Adopts Continuous Integration/Continuous Deployment (CI/CD) for software development/delivery; Modern DevOps – auto-scaling, monitoring feedback loop Software abstraction from the platform Portability; Raises abstraction level from hardware to software components. By consequence,  the Issue, Change and Dependency management environments for the SKAO software were initially adopted as below:

\begin{itemize}
    \item Issue management: Atlassian Jira\texttrademark\cite{Jira}
    \item Change management: ie Subversion (SVN)\cite{Subversion}
    \item Dependency Management: ie., Apache\cite{Apache}, Nexus\cite{Nexus}
\end{itemize}

Although SVN was initially adopted, the standardization of GitLAb made its way and this tool quickly became the open-source version control system and a cornerstone of SKAO DevOps change management.  The Container Orchestration at SKAO supporting cloud infrastructure follows a common containerization framework with portability as a key attribute with key items:  Resource abstraction; Scheduling \& Deployment; Scaling \& Load balancing; Monitoring, Auto-healing \& resource allocation management. A container is essentially a fully packaged and portable computing environment and as such it allows great flexibility while encapsulating and isolating the application with everything an application needs for  proper execution:  binaries, libraries, configuration files, and dependencies. The container itself is abstracted away from the host OS, with only limited access to underlying resources—much like a lightweight virtual machine (VM). As a result, the containerized application can be run on various types of infrastructure—on bare metal, within VMs, and in the cloud without needing to refactor it for each environment. Container Orchestration is the mechanism by which large scale deployment of containers and dependent resources (network,storage,GPU, etc.) is managed and scaled across a multi-node and multi-data centre managed infrastructure environment. This is done seamlessly so that the combined infrastructure can be considered a single resource. The Cloud Native Monitoring \& Control layer follows the following description:

\begin{itemize}
    \item Tango-controls
    \item Hierarchical control structure translates to Kubernetes Resources
    \item Manage deployments with Nested Helm Charts
    \item Inject and inherit configuration with runtime \tt{values.yaml}
\end{itemize}

\begin{figure}
\begin{center}
\begin{tabular}{c}
\includegraphics[height=8cm]{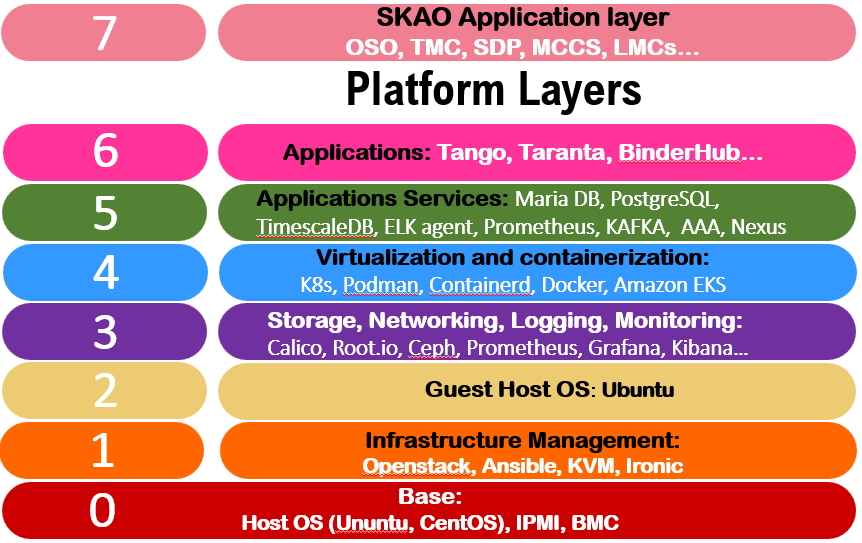}
\end{tabular}
\end{center}
\caption 
{ \label{fig:platform_APIS}
Platform layers and software APIs, from Baremetal to Raw Virtual Machines (VMs) to 
Cloud resources.} 
\end{figure} 

To implement a containerized environment, one needs to run container orchestration to automate the deployment, management, scaling, and networking of containers. Such an orchestrator helps the deployment of an application across different environments without needing to redesign it. Managing the lifecycle of containers with orchestration also supports DevOps (a combination of software development (Dev) and IT operations (Ops)) teams who integrate it into CI/CD workflows. Along with application programming interfaces (APIs) and DevOps teams, containerized applications or services pave the ground for cloud-native applications. For example, container orchestration automate the management of Provisioning and deployment, Configuration and scheduling, Resource allocation, Container availability , Scaling or removing containers based on balancing workloads across your infrastructure, Load balancing and traffic routing , Monitoring container health and security. 

 Containers offer a compelling solution for isolating OpenStack services, but running the control plane on an orchestrator such as Kubernetes or Docker Swarm adds significant complexity and operational overheads. Openstack Kayobe\cite{Kayobe} was adopted as the workhorse version for deployment of containerised OpenStack to bare metal.

The advantages of Containerization include: encapsulation of dependencies; \ac{oci} Image is immutable - provides consistent snapshot; Addresses the well known syndrome of {\it "it worked in Dev - it’s an Ops problem now!"}; Low cost sharing of reliable build/deployment artefacts; Just enough isolation at runtime to enable lightweight multi-tenancy; enables On-board developers and contributors to rapidly Implement reliable CI easily since it is faster and cheaper than creating VMs each time; scales from the desktop to the server.

Each discrete software component, such as a database, device server, or web application, is encapsulated in an OCI container. The OCI specification defines an API protocol to facilitate and standardize the distribution of content, as well as a runtime contract for the definition and execution of an encapsulated application. This specification was launched in April 2018 to standardize container image distribution around the specification for the Docker Registry HTTP API V2 protocol, which supports the pushing and pulling of container images. The runtime parameterization, resource requirements, and application topology are captured in declarative Helm Charts, and scheduled on K8s.
 
Also, for data interchange, the Resource Description Framework (RDF) was adopted: it is a standard model for data interchange on the Web originally designed as a data model for metadata  and support library catalogs and worldwide directories to Mozilla internal data structures or knowledge bases for artificial intelligence projects\cite{books/daglib/0009425}. RDF extends the linking structure of the Web to use URLs to name the relationship between things as well as the two ends of the link (this is usually referred to as a “triple”). Using this simple model, it allows structured and semi-structured data to be mixed, exposed, and shared across different applications. SKAO Agile teams perform transformation from proprietary data (e.g. JIRA issue entries) to standard RDF.

On the other hand, the production and exploitation of industrial control systems that are applicable to the thousands of devices that will operate and monitor SKAO differs substantially from traditional information systems; this is in part due to constraints on the availability and changes in the life cycle of production systems, as well as their reliance on proprietary protocols and software packages with little support for open development standards. The application of agile software development methods therefore represents a challenge which requires the adoption of existing change and build management tools and approaches that can help bridge the gap and reap the benefits of managed development when dealing with industrial control systems. For instance, agile development tools such as Apache Maven for build management, Hudson for continuous integration or Sonatype Nexus for the operation of "definite media libraries" were leveraged to manage the development lifecyle of systems similar to the CERN UAB framework \cite{Unicode,hoffmann:pcapac08-wep007}. 

The SKAO has its observatory software tree structured in several Agile Array Release Trains (ARTs), namely the Observation Management and Control (OMC), the Data processor (DP) ART, the Services ART, MID Integration and Low Integration ARTs. Here, the OMC is meant to address proposal submission \& management, design, scheduling, and execution of observations on both MID and LOW telescopes; correctly operate and monitor the observatory and the telescopes (including Local Monitoring and Control and the Central Signal Processor Local Monitoring and Control ); guide the development of TANGO device interfaces; guide the Integration of engineering operations' software products. 
The Services ART is meant to provide and develop the Compute Platforms, the Networks and the software services to the other ARTs, namely the OMC, MID and LOW Integration. These include: Elasticstack search infrastructure, deployment and management of the Observatory Science Operations Data Archive (ODA), the Observatory Engineering Data Archive (EDA), CI/CD infrastructure (built on GitLab), Jira, Confluence, SKAO email support to Slack, Miro, Google, Platform monitoring, Developer reference implementations, tooling and documentation.
Within the Service ART, ENGAGE SKA has supported the System team, a specialized Agile Team that assists in building and using the Agile development environment, including CI/CD and test automation. The System Team supports the integration of assets from SKAO Agile teams, performs end-to-end solution testing when necessary, and assists with deployment and release. The SKAO implemented one single System Team right before construction, from the bridging phase. This team is responsible for setting the basis and enable the project wide automations needed to implement CI/CD and DevOps. This work is based on the harmonization of previous experiences and practices, as emerged in the pre-construction phase from the experience developed by different consortia. Up to and during construction, the team also contributes to tasks related to system Critical Design Review (CDR) artefacts refinement.  The system Team acted in fact as an operational Agile analogue to a Change Control Board (CCB) for all the software modules version control by the other Agile teams in SKAO and articulated with the Feature and Product Owners of the several ART teams on software Quality Assurance matters.

Over the years, major astronomical observatories like the Atacama Large Millimeter / submillimeter Array (ALMA)\cite{5136193}, the biggest operating millimeter wave interferometer to date, the Vera Rubin Observatory\cite{2016SPIE.9906E..0LA} soon to become the largest wide sky survey machine in the optical domain, the LOw Frequency ARray (LOFAR)\cite{2013A&A...556A...2V} and the Giant Metrewave Radio Telescope (GMRT)\cite{2017CSci..113..707G} currently the two largest long radio wave interferometers in operation), and the four SKA precursors : MeerKAT\cite{10.1117/12.2231602} and the Hydrogen Epoch of Reionization Array (HERA)\cite{DeBoer_2017}, in South Africa, and the two radio telescopes in Western Australia, the Australian SKA Pathfinder (ASKAP)\cite{2010SPIE.7740E..1JG} and the Murchison Widefield Array (MWA)\cite{2013PASA...30....7T}, besides many other large scale observatories, have developed their own customized solutions for their software lifecycle tackling a variety of conditions imposed by their sheer size and often by the nature of their multinational and distributed collaborations. Although SKAO has chosen state of the art trends from the Cloud native world, the ART teams took into account a variety of lessons learned from these projects on the design and decision making process associated with cyberinfrastructure and software lifecycles.

\section{The SKAO software integration environment}
\subsection{the SKAMPI : a tool set concept for Agile delivery}

The SKA Minimal Viable Product (MVP) Integration, or SKAMPI, and as the name implies, integrated every component developed for the SKAO in order to validate their correctness while delivering value in the form of a working prototype. SKAMPI was the conceptual base from which an end-to-end MVP was delivered as self standing artefacts, and functioned as a staging environment where the integration consolidated developed software was run before delivery and connected to real telescope hardware. SKAMPI was both the set of software artefacts, and the corresponding repository and continuous integration facilities that allowed for the development, testing, and integration of the initial SKA prototype software systems, both for SKA Mid and SKA Low telescopes. 

Within the framework of agile development and delivery processes from the ART teams, the SKA used SKAMPI as a vehicle for the automation or deploying and testing systems composed of discretely developed software components, that come together to form the control system for the real-time management of observation sessions, as well as the subsequent processing of observation data. Each component such as the \ac{tm}, the \ac{csp} and the Software Data Processor \ac{sdp}, is developed in accordance with an interface specification created by independent teams, and extensively reviewed and validated.  SKAMPI enabled this distributed development by providing a low cost and automated solution to Continuous Integration and testing (CI/CD).

At this stage, the artefacts delivered should be seen as fully fledged systems (software running on integrated platform servers) capable of providing end-to-end value to the multiple stakeholders. All the tests that are included with SKAMPI were therefore executed with the objective of verifying and validating end-to-end capability.
How those artefacts are assembled, configured and tested is determined by the source code defined within the SKAMPI repository. The software included in every version of SKAMPI consists of a software base with common parts, addressing the SKA Mid, SKA Low and SKA Common/HQ  software components.

\subsection{The SKAMPI Infrastructure}
\label{sect:infra}

SKAMPI relies on a series of modern development practices and tooling to achieve this: Containerisation, Container Orchestration, \ac{bdd} Test Automation, and Continuous Integration (through GitLab).
Containerization is a form of virtualization where applications run in isolated user spaces, ie containers, while using the same shared operating system (OS). \ac{bdd} test suites are defined using the Gherkin language for describing the required business scenarios, which are then arranged into test suites performed against the \ac{skampi} deployment. The entire process is automated using GitLab CI, with the test results being captured and fed back to the developers on a continuous basis so that issues and progress can be tracked and understood. 

\begin{figure}
\begin{center}
\begin{tabular}{c}
\includegraphics[height=10cm]{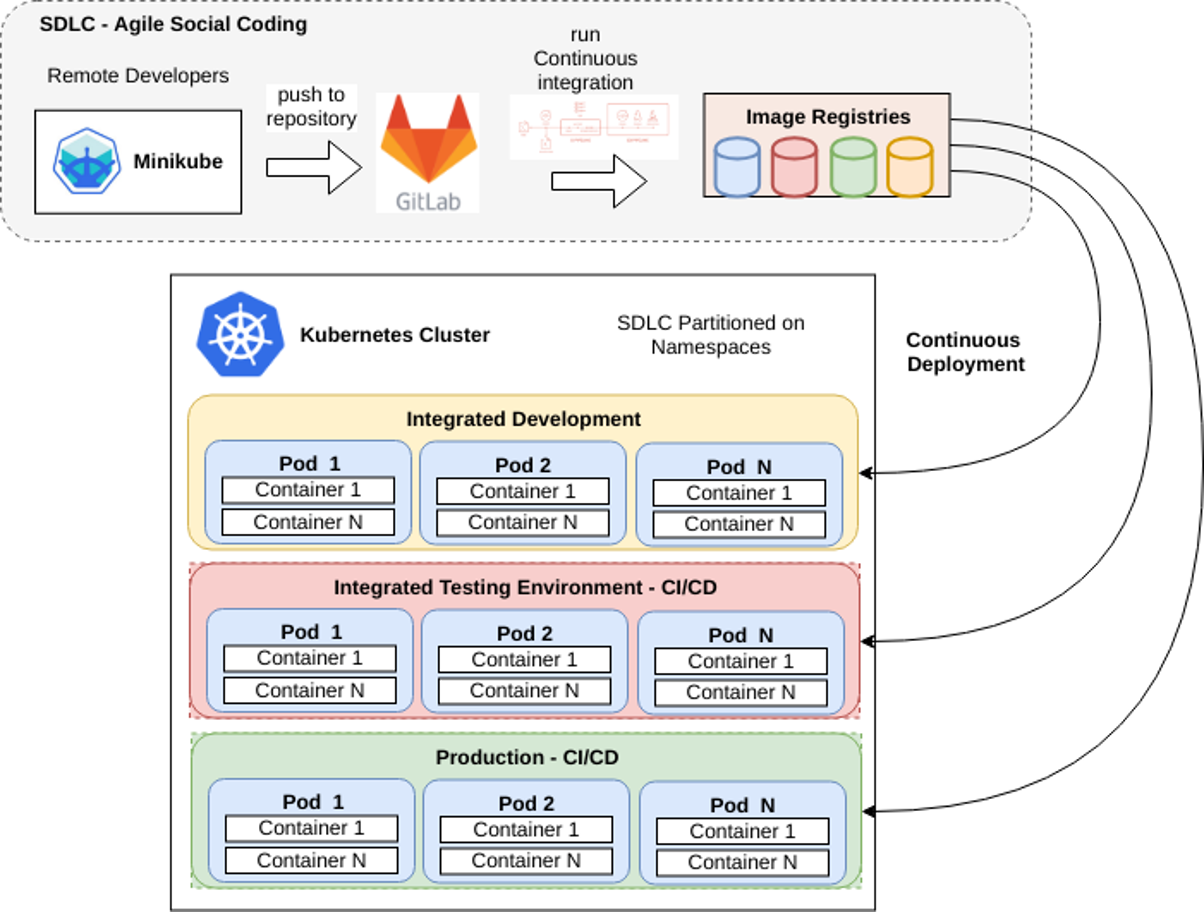}
\end{tabular}
\end{center}
\caption 
{ \label{fig:K8s_helm_namespace}
The SKAO Agile Software Development Life Cycle (SDLC).} 
\end{figure} 

SKAMPI suffered from a lack of ownership problems that magnified the brittle nature of that environment. The OMC Array Release Train (ART) has several components and moving parts that could behave in unexpected ways before being fully tested and properly configured. That ended up causing outages in SKAMPI that took dedicated personnel to constantly fix them.

While SKAMPI has since been deprecated, the fundamental ideas behind it live on as the \ac{ska} project has moved on into an Array Integration and Verification (AIV) phase. The fundamentals have been split out into two repositories that mirror the requirements for the two telescopes (Mid and Low), and enable software teams to deliver independent releases of components at the product level. This addition to the philosophy behind SKAMPI has enabled a degree in improvement of decoupling software delivery between components making it easier to deliver continuous integration. This is predicated on the development of robust component interface definitions (ICDs) and the implementation of interface deprecation and compatibility protocols as well as feature flagging.

\subsection{EngageSKA Cluster}
\label{sect:infra:cluster}
To support development activities of the different \ac{ska} SAFe teams that included SKAMPI integration and support to the System team (Services ART),  in particular during the Bridging Period towards SKA formal construction and early construction phase, the software development and validation lifecycle was based on a purposely built cluster, located at ENGAGE Research Infrastructure, a native cloud based platform located at Institute of Telecommunications in Aveiro, Portugal. ENGAGE SKA cloud compute node took its name from the ENGAGE SKA Research Infrastructure of the Portuguese Roadmap of Research Infrastructures of Strategic Relevance that federated efforts across Portugal's bid to become a member of the SKA Observatory. This cluster not only supported the development efforts of the teams, but it was also used to run and test SKAMPI, including each one of its components that result from the global development effort. Additionally, several other clusters in different location (STFC Cloud, PSI Low, PSI Mid, ITF Low, Low CSP, ITF Mid) were forwarded to enhance performance, centralize processes, provide scalable resources therefore providing the necessary infrastructure redundancy and following the principles here outlined. At later stages, the STFC Iris platform became the natural cloud backbone for SKA software development and support.

EngageSKA provided an invaluable proving ground for testing and iterating over the design, tools, techniques and Software Defined Infrastructure (SDI) recipes that have since transformed into the model for the current iteration of the Array Assemblies (AA) AA0.5/AA1 production platforms (Mid and Low) for the \ac{ska}. The pre-construction availability of the EngageSKA platform created an ideal space for nurturing and developing the skills required in the core Systems and Platform Team of the \ac{ska} needed to support the dynamic and demanding platform and infrastructure requirements of AIV and production.

The physical cluster was comprised of 16 Rack servers, specked as detailed in Table \ref{tab:engageska_servers}. These servers have specific roles, ranging from compute, controller, storage and GPU accelerators. The \emph{Compute A} servers were the first set of servers acquired to support the EngageSKA cluster, followed by the \emph{Compute B} and then the \emph{Compute C} servers at later dates. This acquisition offset in time is the reason why the computational resources available increase from one set of servers to the next. The \ac{gpu} node is equipped with an NVIDIA Tesla V100-PCIE-32GB graphics processing unit to support any development needs that require acceleration using specialized hardware and CUDA. The Storage nodes are used to store any and all data related to the development process, component integration and testing monitoring, etc. They provide several terabytes of storage, with specific performance characteristics. Some are supported by rotational media, while others are entirely comprised of fast \ac{ssd} media. Finally, the \emph{Controller} nodes orchestrate the virtual machines of the various teams running of the compute nodes and their data that is stored on the Storage nodes.  

\begin{table}[ht]
\caption{EngageSKA servers.} 
\label{tab:engageska_servers}
\begin{center}       
\begin{tabular}{c|c|c|c|c} 
\hline
\rule[-1ex]{0pt}{3.5ex}  Server Type & Quantity & \#vCPU & Memory (GB) & Storage (GB)\\
\hline
\rule[-1ex]{0pt}{3.5ex}  Compute A & 5 & 48 & 128 & 446,6 \\
\rule[-1ex]{0pt}{3.5ex}  Compute B & 3 & 48 & 512 & 446,6 \\
\rule[-1ex]{0pt}{3.5ex}  Compute C & 2 & 128 & 512 & 1117,25 \\
\rule[-1ex]{0pt}{3.5ex}  GPU & 1 & 32 & 128 & 200+12315 \\
\rule[-1ex]{0pt}{3.5ex}  Controller & 2 & 16 & 64 & 1798,7 \\
\rule[-1ex]{0pt}{3.5ex}  Storage & 3 & 16 & 64 & 200+12315 \\
\hline 
\end{tabular}
\end{center}
\end{table} 

\begin{figure}
\begin{center}
\begin{tabular}{c}
\includegraphics[height=10cm]{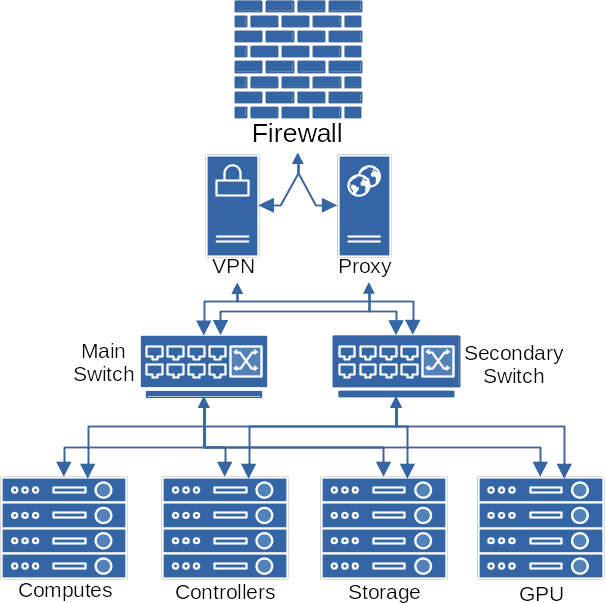}
\end{tabular}
\end{center}
\caption 
{ \label{fig:engageska_cluster_diagram}
EngageSKA Cluster Diagram. } 
\end{figure}

These nodes were also used to support other projects related to \ac{ska}, such as the SKA Data Challenges that were brought up as exercises aiming to inform the development of the data reduction pipelines for the Software Data Processor (SDP) and the SRCnet nodes, to allow the science community to get familiar with the standard products the SKA telescopes will deliver, and optimise their analyses to extract science results from them. Within these Data Challenges, multiple teams compete over the course of multiple months to process astronomical data. 
Fig. \ref{fig:engageska_cluster_diagram} shows a cluster diagram indicating how it is set up.

Accessing the cluster by distributed users can be done in one of two ways: through a Proxy Server that listens on specific ports to expose selected services to the internet, or through a \ac{vpn} server that allows access to specific segments of the internal network. The EngageSKA servers are connected to the internal network through two redundant switches that provide reliable communication, with protection against a single point of failure. Likewise, and as expected, all servers have redundant power supplies, redundant cabling, redundant \ac{ups} and a backup Diesel generator. 

\subsection{Architecture}
\label{sect:infra:arch}
\begin{figure}
\begin{center}
\begin{tabular}{c}
\includegraphics[height=10cm]{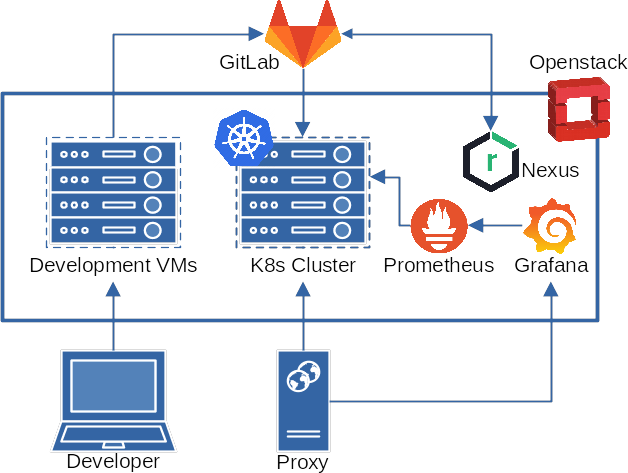}
\end{tabular}
\end{center}
\caption 
{ \label{fig:ska_arch_diagram}
SKA Software Development Architecture. } 
\end{figure}

The software development for \ac{ska} is supported on top of the EngageSKA cluster, which is shown in Fig. \ref{fig:ska_arch_diagram}. The cluster hosts an OpenStack\cite{Openstack} deployment, which hosts all the development virtual machines and a K8s cluster called \verb|syscore|.  

During system upgrades, it was found that thin-provisioning the images saved a lot of migration time and enabled dynamical memory allocation (for instance passing From 100GB to ~3GB thanks to \verb|libguestfs| for the \verb|virt-sparsify| tools). \verb|libguestfs| is a set of tools for accessing and modifying virtual machine (VM) disk images while \verb|virt-sparsify| can make virtual machine disks (or any disk image) sparse (or thin-provisioned). This means that free space within the disk image can be converted back to free space on the host, therefore enabling a much robust compute resource management.

\verb|Virt-sparsify| can locate and sparsify free space in most filesystems (eg. ext2/3/4, btrfs, NTFS, etc.), and also in LVM physical volumes.

\verb|Virt-sparsify| can also convert between some disk formats, for example converting a raw disk image to a thin-provisioned \verb|qcow2| image. \verb|Virt-sparsify| can operate on any disk image, not just ones from virtual machines. However if a virtual machine has multiple disks and uses volume management, then \verb|virt-sparsify| will work but not be very effective.

The developers connect to their own development machines using a VPN connection, and after they are finished with a piece of software this is normally pushed into GitLab repositories. The repositories are setup with continuous integration and deployment pipelines, which run on the \verb|syscore| K8s cluster using GitLab runners. These pipelines can also deploy new components for SKAMPI, which was updated on the integration namespace every day. Once \ac{skampi} was deployed on \verb|syscore|, the proxy server allows to access its user interface to control a telescope mocked by its different components.

The Openstack cluster also hosts other services, such as: a Nexus repository manager where the different components are versioned and can be accessed during development; an Elasticsearch\cite{Elasticsearch} cluster to collect and index the logs; a Kibana application to show and query the log data; a Prometheus\cite{Prometheus} server for monitoring; and Grafana dashboards to display the information provided by Prometheus that can be accessed through a proxy server from the internet. Another DevOps related integration not showed in the diagram is the Slack platform , which is used as a secure messaging platform inside SKAO and, more precisely, it receives alert messages of any issues that arose on the SKAMPI deployments or with any of the cluster nodes. Based on this information, developers and product owners decide which improvements must be made and create tickets on the issue tracking platform used on the organization, JIRA.

The selected logging solution is Elasticsearch, storage is handled through Ceph\cite{Ceph}, while Prometheus handles monitoring (see the Prometheus section below), and the (central) artefact repository (CAR) is Nexus. It is important to realize that only artefacts produced by GitLab pipelines that have been marked for a release (i.e. triggered by a git tag), are allowed to be stored on the CAR. On all other cases, GitLab’s own artefact repository is used. ENGAGE SKA has selected Prometheus for evaluation as monitoring solution for the SKA CICD (continuous integration and continuous deployment) Infrastructure that became later adopted with the help of Thanos by SKAO as the current CI/CD monitoring solution\cite{dicarlo:icalepcs2023-tupdp045}. 

\section{Cluster tools for developers}
\label{sect:user}

\subsection{Tango}
\label{sect:user:tango}
Tango Controls\cite{Tango-controls} is a free open source device-oriented controls toolkit for controlling any kind of hardware or software and building SCADA (supervisory control and data acquisition) systems and nowadays has a very rich ecosystem that includes all these application. 

The SKAO  has selected TANGO-controls as the distributed middleware for driving software and hardware devices using the Common Object Request Broker Architecture (CORBA)\cite{CORBA} distributed objects that represent devices that communicate with ZeroMQ\cite{ZeroMQ} events internally. The entire system runs on a containerised environment managed by Kubernetes (K8s) with Helm\cite{Helm} for packaging and deploying SKA software. As described in \cite{dicarlo:icalepcs2023-tupdp045}, K8s uses a declarative “model” of operation that drives the system towards the desired state described by user manifests, with various controller components managing the lifecycle of the associated resources. Helm provides the concept of a chart which is a recipe to deploy multiple K8s resources (i.e., containers, storage, networking components, etc...) required for an application to run. The resources are created using templates so that the chart can adapt generic configurations to different environments (i.e., the different SKA datacentres). The SKA deployment practices include the heavy use of standardised Makefiles (i.e., for building container images, for testing, for the deployment of a chart, etc.) and GitLab for the CICD (continuous integration continuous deployment) practices.

Tango Controls is Operating System (OS) independent and supports C++, Java and Python for all its components. Tango Controls has been enriched with many applications (desktop and web based) that can satisfy almost all the user needs since its is a device-oriented controls toolkit, can write custom applications but can also work as a final product. Among its applications, we can find real-time alarm with alarm status, historical and real-time data monitoring. Besides its reals time control messaging of up to several dozens of thousands of devices Tango Controls allows for forensic  analysis of engineering and operating centre databases in particular. Indeed, Tango benefits from a strong Community of Practice (CoP) that benefits large scale distributed sensor machines like high energy physics instruments (ie accelerators and synchrotrons) and radioastronomy projects.

\begin{figure}
\begin{center}
\begin{tabular}{c}
\includegraphics[width=15cm]{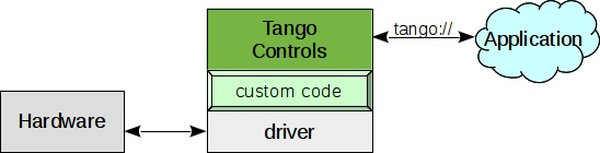}
\end{tabular}
\end{center}
\caption 
{ \label{fig:what_is_tango}
What is tango? } 
\end{figure}

\section{Pitfalls and Life Lessons}
\label{sect:pitfalls}
Perhaps the most challenging case within ENGAGE SKA cluster management was the planing of resource upgrades aiming for an automated deployment of the cyberinsfracture and the associated software stack. The previous EngageSKA cluster had resource management issues: with time passing, many instances were without ownership information, with large volumes attached (upwards to $\sim1000$GB) and consuming valuable resources (up to 8 CPUs and 16GB RAM). Key points from deployment experience: i)the use of a centralized configuration and automated deployment and monitoring, benefiting from Ansible-based solutions. ii) Converting volumes to images turned instance migration much easier with \verb|libguestfs|.

During the Openstack migration and rebuild of the software stack, thin-provisioning the images saved a lot of migration time (In our case from $\sim100$GB to $\sim3$GB, after choosing \verb|libguestfs| tool suite. As described above, SKAMPI suffered from a lack of ownership problems that magnified the brittle nature of that environment. The OMC Array Release Train (ART) has several components and moving parts that could behave in unexpected ways before being fully tested and properly configured. That ended up causing outages in \ac{skampi} that took dedicated personnel to constantly fix them.

More important, basic containerisation approach for artefact delivery future proofs us (somewhat) and adhering to language specific frameworks, tooling and standards actually helps to reduce what must be maintained and increases portability. This implied a great effort across the collaboration to concentrate on {\it "keep it simple and tight"}, publishing standards, reference implementations and providing supporting resources. In the end, the SDLC conditions are such it should enable  continuous guidance and tooling for upgrading. At the API level, efforts have been developed in providing to the user/developer standardized API as much as possible: API vanilla (industry standard) and small (eg: support core native resource types). This keeps a clear separation between API and platform implementation and helps the Agile teams to resist the urge for anything to be special.

\section{Conclusions} We outlined the major characteristics and innovation approaches to address the cloud platform architecture and software life cycle for the SKAO OMC and Services ARTs using the ENGAGE SKA cloud platform. ENGAGE SKA platforms have been instrumental to provide a development path towards long-term sustainability and maturity in the implementation of the SAFe\textsuperscript{\textregistered} Framework to help SKA become a Lean Enterprise and achieve Business Agility. 

ENGAGE SKA cluster has supported an outstanding community around it, keeping solution development and deployment an active and feature-rich process. Within the SKA software Services ART, ENGAGE SKA has supported the System team, a specialized Agile Team that assists in building and using the Agile development environment, including CI/CD and test automation. The System Team supported the integration of assets from SKAO Agile teams, performed end-to-end solution testing when necessary, and assisted with software deployment and release. 

Cloud native environment provides a target environment and framework for developers to aim for and enables a greater possible portability from desktop to production. Additionally, it insulates activity from production platform concerns with the latest possible practical hardware commitment, lowers barrier to entry, reduces time-to-share and enables integration and testing in a completely virtual environment by providing an abstraction from the underlying infrastructure.

Monitoring the performance of the infrastructure put in place for CI/CD purposes relied on Prometheus that was adopted by SKAO together with with Thanos and Grafana for the CI/CD scope. As described, the main rationale for this choice is the modularity of the several tools within the Prometheus ecosystem (custom exporters, data-sources and visualizations), together with performance efficiency (optimal usage of memory allocation) and scalability (vertical only at the moment). 

\acknowledgments

This research was supported by the project Enabling Green E-science for the SKA Research Infrastructure (ENGAGE SKA), reference POCI-01-0145-FEDER-022217, funded by COMPETE 2020 and FCT, Portugal and Project Centro de Investigação em Ciências Geo-Espaciais, reference UIDB/00190/2020, funded by COMPETE 2020 and FCT, Portugal; JPB acknowledges support from Project Lab. Associado UID/EEA/50008/2019. MdC has been supported by the Italian Government (MEF- Ministero dell’Economia e delle Finanze, MIUR - Ministero dell’Istruzione, dell’Università e della Ricerca). The ENGAGE SKA team acknowledges all the support and collaboration from the SKA Observatory. DN acknowledges support from the HPC Chair at University of Évora.





\end{document}